# Coupling of a Single Tin-vacancy Center to a Photonic Crystal Cavity in Diamond

Kazuhiro Kuruma[1,a)], Benjamin Pingault[1,2], Cleaven Chia[1], Dylan Renaud[1], Patrick Hoffmann[3], Satoshi Iwamoto[4], Carsten Ronning[3], and Marko Lončar[1,a)]

[1]*John A. Paulson School of Engineering and Applied Sciences, Harvard University, Cambridge, MA 02138, USA*

[2]*QuTech, Delft University of Technology, PO Box 5046, 2600 GA Delft, The Netherlands, EU*

[3]*Institute of Solid State Physics, Friedrich Schiller University of Jena, Max-Wien-Platz 1, 07743, Jena, Germany*

[4]*Research Center for Advanced Science and Technology, The University of Tokyo, 4-6-1 Komaba, Meguro-ku, Tokyo 153-8505, Japan*

a) kkuruma@seas.harvard.edu and loncar@seas.harvard.edu

**Abstract**

**We demonstrate optical coupling between a single tin-vacancy (SnV) center in diamond and a free-standing photonic crystal nanobeam cavity. The cavities are fabricated using quasi-isotropic etching and feature experimentally measured quality factors as high as ~11,000. We investigate the dependence of a single SnV center's emission by controlling the cavity wavelength using a laser-induced gas desorption technique. Under resonance conditions, we observe an intensity enhancement of the SnV emission by a factor of 12 and a 16-fold reduction of the SnV lifetime. Based on the large enhancement of the SnV emission rate inside the cavity, we estimate the Purcell factor for the SnV zero-phonon line to be 37 and the coupling efficiency of the SnV center to the cavity, the $\beta$ factor, to be 95%. Our work paves the way for the realization of quantum photonic devices and systems based on efficient photonic interfaces using the SnV color center in diamond.**

Color centers in diamond are promising solid state quantum emitters, of interest for realization of single photon sources [1] and quantum memories leveraging their long-lived spins [2]. The nitrogen-vacancy (NV) center, the most intensively investigated among diamond color centers, has been used in quantum network demonstrations, including photon-mediated remote entanglement of distinct NV centers [3] and deterministic delivery of entanglement between nodes of a quantum network [4,5]. However, NV centers have a low zero-phonon line (ZPL) emission (only ~3% of total emission), and tend to be unstable (blinking, spectral diffusion, etc) inside nanostructures, due to a high susceptibility to external electric field fluctuations from surfaces[6]. Recently, group-IV color centers in diamond, such as the silicon-vacancy (SiV) [7–9], germanium-vacancy (GeV) [10–12], and tin-vacancy (SnV) [13–15], have attracted much attention because of their stable and large ZPL emission, even inside nanostructures. While the SiV and GeV centers require operation at mK temperature [16] or under static strain [17] to suppress spin decoherence caused by phonon-induced transitions between ground state levels, the SnV centers have a much larger splitting between these levels (~850 GHz), and can thus support long spin coherence times at 1 K [15].

To take advantage of the full potential of SnV centers, it is important to realize an efficient spin-photon interface for SnV that can improve the ZPL collection efficiency and enhance the coherent ZPL emission by resonant coupling to color centers [18–21]. This can be accomplished using optical micro/nanocavities with high quality ($Q$) factors. Photonic crystal (PhC) nanocavities are particularly promising platforms for enhancing the light-matter interaction, owing to their high $Q$ factors and small mode volumes ($V$). In particular, 1-dimensional (1D) nanobeam cavities have been intensively studied in diamond because they can be readily fabricated using a variety of techniques, including angle etching [22], diamond film thinning [23,24], and quasi-isotropic etching [25]. Importantly, high-$Q$ 1D nanobeam cavities coupled with SiV centers have recently enabled the experimental demonstration of a single photon transistor [9] and spin memory-enhanced quantum communication [26]. Finally, nanobeam cavities can also support high-frequency (~ 10 GHz) and high-$Q$ mechanical modes [27], which are of interest for the realization of spin-phonon interfaces.

In this letter, we report on optical coupling of a single SnV center to a PhC nanobeam cavity in diamond. We use a quasi-isotropic undercut method to fabricate free-standing PhC nanobeam cavities in bulk diamond crystal, featuring $Q$ factors as high as 11,000. Next, we select a cavity ($Q$ ~ 3,000) with a good coupling to a single SnV center, introduced via ion-implantation, and tune the cavity into resonance with the SnV center using a laser-induced gas desorption technique. At resonance, we observe a strong enhancement of the SnV emission intensity by a factor of 12 and a 16-fold reduction in the SnV lifetime (compared to centers in bulk). Due to the significant enhancement of the spontaneous emission rate, we estimate a Purcell factor for the SnV ZPL of 37 and a near-unity cavity coupling efficiency $β$ (probability of emitted photons being channeled into the cavity mode) of 95%. These results provide a step towards the development of quantum information processing devices including efficient quantum light sources [23,28], and the creation of spin-photon interfaces [26] using single SnV centers coupled to high-$Q$ PhC cavities.

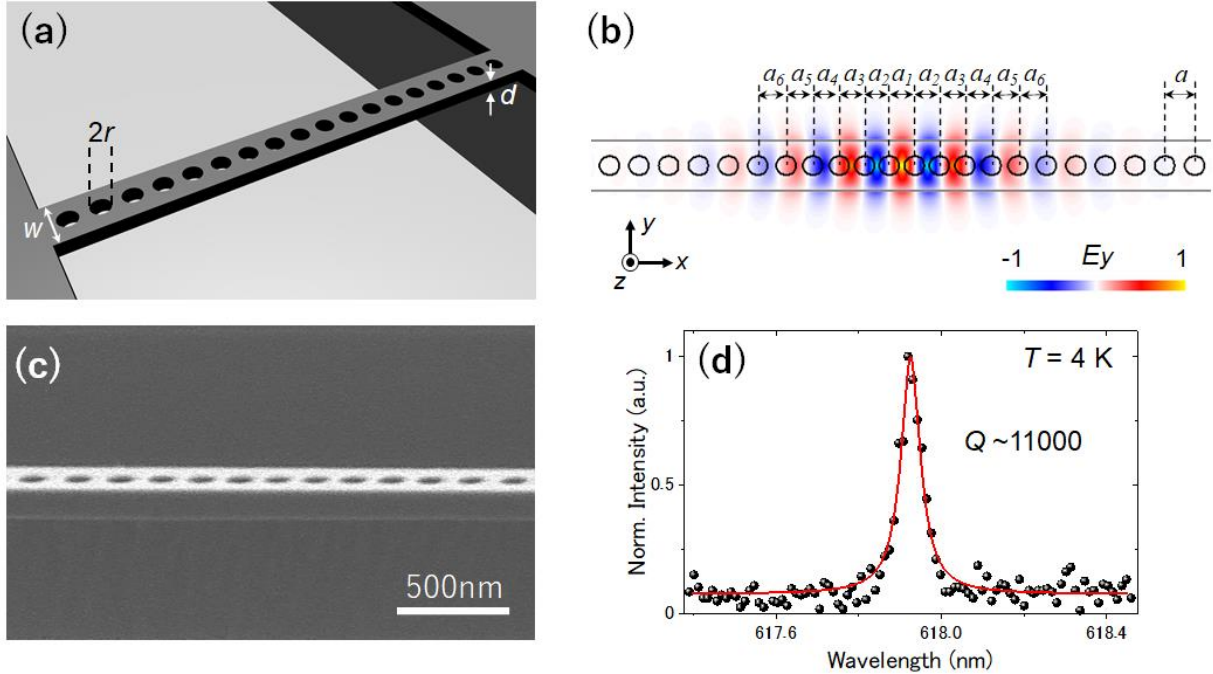

**Fig. 1.** (a) Schematic view of the free-standing 1D PhC nanobeam cavity on bulk single-crystal diamond. (b) Calculated electric field distribution for the fundamental cavity mode with a theoretical $Q$ factor of ~$6\times10^5$. The quadratic modulation of the air hole positions ($a_1$~$a_6$) is applied near the center of the waveguide to form the cavity. (c) Scanning electron microscope image of a fabricated cavity. (d) Selected PL spectrum of a cavity with $a$ =188 nm, exhibiting the highest $Q$ factor among the 153 measured cavities. The red solid line is a fitting curve.

We used a diamond based PhC nanobeam cavity, as schematically shown in Fig 1(a). The cavity structure consists of a free-standing nanobeam waveguide with a width ($w$) of 330 nm and thickness ($d$) of 170 nm. The lattice constant ($a$) ranges from 188 to 196 nm, and the airhole radius ($r$) is 65 nm. In order to form the cavity, the air holes are shifted near the waveguide center quadratically [29]($a_1 = 0.84a$, $a_2 = 0.844a$, $a_3 = 0.858a$, $a_4 = 0.88a$, $a_5 = 0.911a$, and $a_6 = 0.951a$). For $a = 196$ nm, we obtained a high theoretical $Q$ factor of ~$6\times10^5$ and a very small mode volume $V$ of ~$0.42(\lambda/n)^3$ for the fundamental cavity mode by a 3D finite-difference time domain method (refractive index of the diamond slab is 2.4). Figure 1(b) shows the cavity design, overlaid with the calculated electric field ($E_y$) distribution of the fundamental cavity mode.

For the fabrication of the designed cavity in diamond, a commercially available electronic grade single-crystal diamond sample (Element Six) was etched using argon/chlorine followed by oxygen plasma in order to remove the surface damaged layer caused by polishing. It was then cleaned by immersion into a 1:1:1 boiling tri-acid mixture of perchloric, nitric, and sulfuric acid for one hour[30]. Tin ($^{117}$Sn) atoms were implanted with an energy of 350keV resulting into a mean ion range of about 86 nm from the surface with a straggling of 17 nm, as simulated by the software package "Stopping and Range of Ions in Matter" (SRIM) [31]. The sample was divided into two areas and implanted with two different ion fluences of $2\times10^{10}$ and $1\times10^{12}$ ions/cm². We employed the lower implantation dose area for optical experiments using a single SnV center later. During the implantation, the sample was tilted by 7° in order to avoid channeling effects. The sample was subsequently tri-acid cleaned again, and then annealed at 1200°C for approximately 5 hours to allow vacancies to become mobile and form SnV complexes with the implanted Sn atoms.

Following another tri-acid cleaning, free-standing PhC nanobeam cavities were fabricated using a combination of electron beam (EB) lithography and dry etching processes. We then deposited a 100nm-thick SiN layer on the bulk diamond by plasma-enhanced chemical vapor deposition. The SiN layer is used as a hard mask for the diamond etch later. We conducted EB lithography to write the cavity structure on a 400nm-thick EB resist (ZEP-520A). After the resist development, we etched the SiN mask layer by induced coupled plasma reactive ion etching (ICP-RIE) with sulfur hexafluoride ($SF_6$) and octafluorocyclobutane ($C_4F_8$) gases. After removing the EB resist, the cavity pattern was transferred onto the diamond substrate by oxygen-plasma RIE. A 20nm-thick $Al_2O_3$ layer was deposited by atomic layer deposition for conformal coverage of the sample and then etched out by RIE, while keeping the sidewalls of the nanobeam to be covered with $Al_2O_3$. In order to realize free-standing cavity structures, we employed a quasi-isotropic undercut technique [25] using oxygen-based RIE. We finally removed the SiN and $Al_2O_3$ layers by immersion in hydrofluoric acid (HF). Figure 1 (c) shows the scanning electron microscope (SEM) image of one of the fabricated nanobeam cavity.

We characterized the fabricated devices using photoluminescence (PL) measurements at 4K, in a closed-cycle cryostat (Montana Instruments). We excited the sample with a 520nm continuous wave diode laser, and the PL signal from the sample was collected through an objective lens (Olympus MPLFLN, 100× magnification and 0.9 NA) and analyzed by a spectrometer (Princeton Instruments) equipped with a Si CCD camera. Figure 1(d) shows a PL spectrum for the fundamental cavity mode with the highest *Q* factor among 153 measured cavities. By fitting the spectrum with a Lorentzian function (red line), the measured *Q* was deduced to be ~11,000, which is comparable to that of visible-wavelength nanobeam cavities realized by quasi-isotropic etching [25].

Next, we investigated the optical coupling between a single SnV center and a nanobeam cavity with $a$ =196 nm. The measured $Q$ factor of the cavity was ~ 3000. The cavity was tuned in and out of resonance with a single SnV ZPL transition at 624nm using the gas tuning method, discussed below. Figure 2 (a) shows the PL spectra taken under various cavity-SnV detunings. The longer wavelength of this SnV compared to the usual 620nm reported in other works [13,14], is due to residual strain [32], induced by the ion implantation and/or the fabrication. Unstrained SnV centers were also present in cavities, but most single ones could not be spectrally isolated. The linewidth of the SnV emission is below the spectral resolution limit of our spectrometer (~10GHz). Note that the SnV emission in the nanobeam cavity did not exhibit any large spectral shifts over our spectrometer resolution through the whole experiments, suggesting the suitability of SnV centers for integration into nanophotonic structures. For the control of the detuning between the SnV and the cavity wavelengths ($\Delta=\lambda_{SnV} - \lambda_{cavity}$), we employed a gas tuning technique based on local heating using focused laser excitation [33]. After injecting the $N_2$ gas in the cryostat chamber, the gas condenses on the cavity resulting into a red-shift of the cavity wavelength over the target SnV emission. Subsequently, we used high excitation powers (> 1 mW) of the CW laser at 520 nm to locally heat the cavity close to its center, which results in the desorption of the $N_2$ gas from the sample and enables a controlled blue-shift of the cavity wavelength. The magnitude of the shift can be controlled by varying the laser power and excitation time. It is noteworthy that we did not see significant changes in the $Q$ factors during the gas tuning. Figure 2 (b) and (c) show PL spectra under off-resonance ($\Delta=$ -1.0 nm) and on-resonance ($\Delta\sim0$ nm) conditions. When the cavity is resonant with the SnV emission, we observed a 12-fold enhancement of the SnV emission, suggesting an increased spontaneous emission rate by the Purcell effect. For the estimation of the enhancement factor, we compared the peak area under the Lorentzian fit curves of the SnV emission in the on- and off-resonance cases. We note that out of 32 cavities we measured, all featuring the same $a$ and typical $Q$ factor >$10^3$, only one could be tuned on- and off-resonance with a spectrally isolated single SnV center. This low yield is mainly because many cavities contain multiple SnV emitters with close or overlapping ZPL emissions, or the cavity-SnV wavelength detuning is large. In order to increase the yield of devices with good SnV-cavity coupling, it may be needed to employ ion implantation with lower doses and lower energies, to reduce the number of centers per cavity and residual strain, as well as employ techniques with precise spectral and spatial alignment of cavities with respect to the single color centers [34].

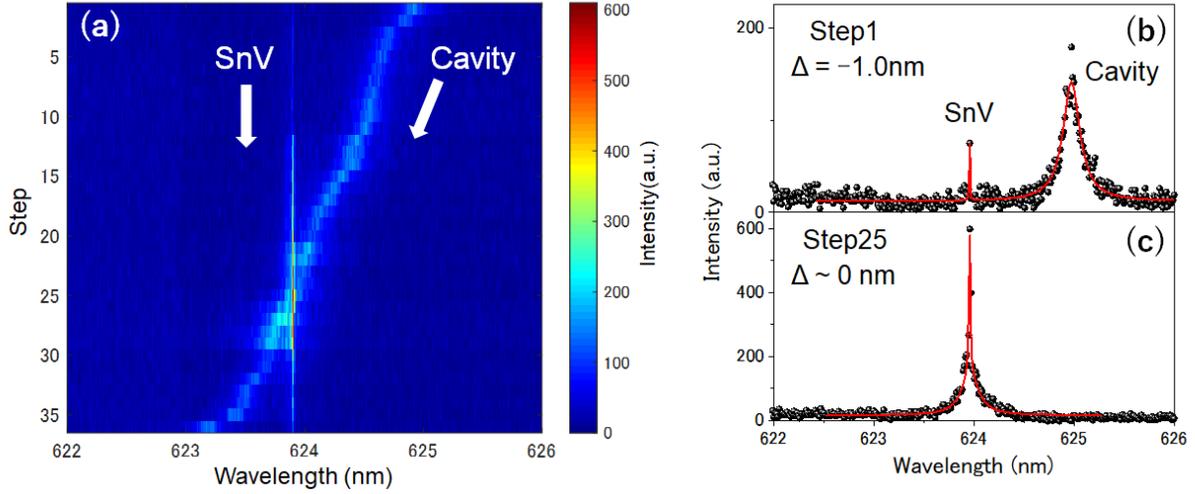

**Fig. 2.** (a) Color map of low-temperature (4 K) PL spectra taken when the cavity resonance is tuned through the SnV emission by a gas desorption process using a focused laser beam. PL spectrum measured under the detuning (b) Δ = -1.0 nm and (c) Δ ~ 0 nm. Red solid lines are fitting curves with multiple Lorentzian peak functions.

We also carried out time-resolved PL measurements on the same SnV for different values of the SnV-cavity detuning. We excited the sample using a supercontinuum laser (SuperK EXTREME) with a central wavelength of 535nm, pulse width of 100nm and a repetition rate of 9.74MHz. The averaged pump power was fixed to 300 $\mu$W. Spectral filtering of the SnV emission was performed using a home-made monochromator. The filtered signal was sent to an avalanche photodiode (APD, Excelitas SPCM) and analyzed by a single photon counting module (PicoHarp 300). The time resolution of our system was measured to be 600ps. Figure 3(a) shows the measured PL decay curves with 5 different detunings of Δ = 0.88nm, 0.38 nm, 0.27 nm, 0.17 nm, and ~0 nm. As expected, the emission lifetime of the SnV is reduced as the detuning becomes smaller. We fitted the PL curves with one or two exponential functions convolved with the gaussian response function with our system's time resolution. The extracted emission rates of the PL curves are plotted in Fig. 3(b), showing a clear enhancement of the spontaneous emission rate of the SnV near resonance. In particular, the fastest lifetime, near zero detuning (on resonance, $\gamma_{on}$), is 0.38 ns, which is approximately 16 times faster than that measured for SnV in bulk (plotted using black dots in Figure 3(a)). We noticed that the PL curve at resonance also exhibits a slow decay with a lifetime of 4.6 ns. This slow decay component mainly originates from the bare cavity emission, which was confirmed by measuring the cavity lifetime under a far-detuned condition to be 4.4 ns.

From the extracted lifetimes, we estimate the Purcell factor of the investigated ZPL ($Fp^{ZPL}$) using the following equation [23]; $Fp^{ZPL} = (\gamma_{bulk} / \gamma_{on} - \gamma_{bulk} / \gamma_{off}) / \xi_{ZPL}$. Here $\gamma_{bulk}$, and $\gamma_{off}$ are the lifetime of SnV in bulk, and off resonance, respectively. $\xi_{ZPL}$ is defined by the fraction of the total emission into the stronger of the two ZPL transitions visible at 4K for SnV in bulk diamond (~ 0.4), which is roughly estimated by a product of the Debye–Waller factor of 57% [35] and the branching ratio of approximately 70% into that transition at 4K [15]. Using $\gamma_{bulk}$ = 6.0 ns, $\gamma_{on}$= 0.38 ns, $\gamma_{off}$ = $\gamma_{0.88nm}$= 7.4 ns, we deduced the Purcell factor to be 37. This value is still smaller than the theoretical Purcell factor [36] of 362 estimated using the measured $Q$ ~ 3,000, and $V$ ~ 0.42 $(\lambda/n)^3$ and considering the polarization mismatch between the dipole moment of the SnV transition <111> and the cavity local electric field <110>. The reason for the reduced Purcell factor could be a sub-optimal position of the SnV center with respect to the antinode of the cavity electric field, due to the stochastic nature of the implantation process. For further improvement of the Purcell factor, the use of precise positioning techniques of color centers, [9,34,37] as well as high $Q/V$ cavity designs [38,39] could be necessary. Using the lifetimes on and off resonances, we also obtain a high coupling efficiency of the SnV transition to cavity mode, $\beta = 1/\gamma_{on}/(1/\gamma_{on} +1/ \gamma_{off}) = $ 95%, indicating a near unity probability for emitted photons from the SnV transition to be channeled into the investigated cavity mode.

Finally, in order to confirm the single photon nature of the investigated SnV emission, we performed second-order correlation measurements for the SnV transition using a Hanbury Brown-Twiss setup equipped with two APDs. We used a laser repetition rate of 77.9MHz with an average pump power of 2.7 mW. Figure 3 (c) shows the intensity correlation histogram measured when the SnV is slightly detuned from the cavity resonance by 0.23nm (see inset). The second-order correlation function at zero delay time, $g^2(0)$, exhibits a clear antibunching with a value of 0.27, confirming that the investigated SnV is a single photon emitter. We consider that the non-zero value of $g^2(0)$ could be primarily due to the contribution of the background cavity emission [40] supplied by other off-resonant SnV centers inside the cavity. In order to reduce the value of $g^2(0)$, it would be necessary to use lower ion implantation doses or deterministic implantation techniques [9,37] to largely reduce the SnV center density inside a single cavity. The $g^2(\tau)$ histogram also exhibits a bunching behavior visible on the first few peaks, which suggests the existence of a shelving state other than the excited and ground states in the SnV [13,41].

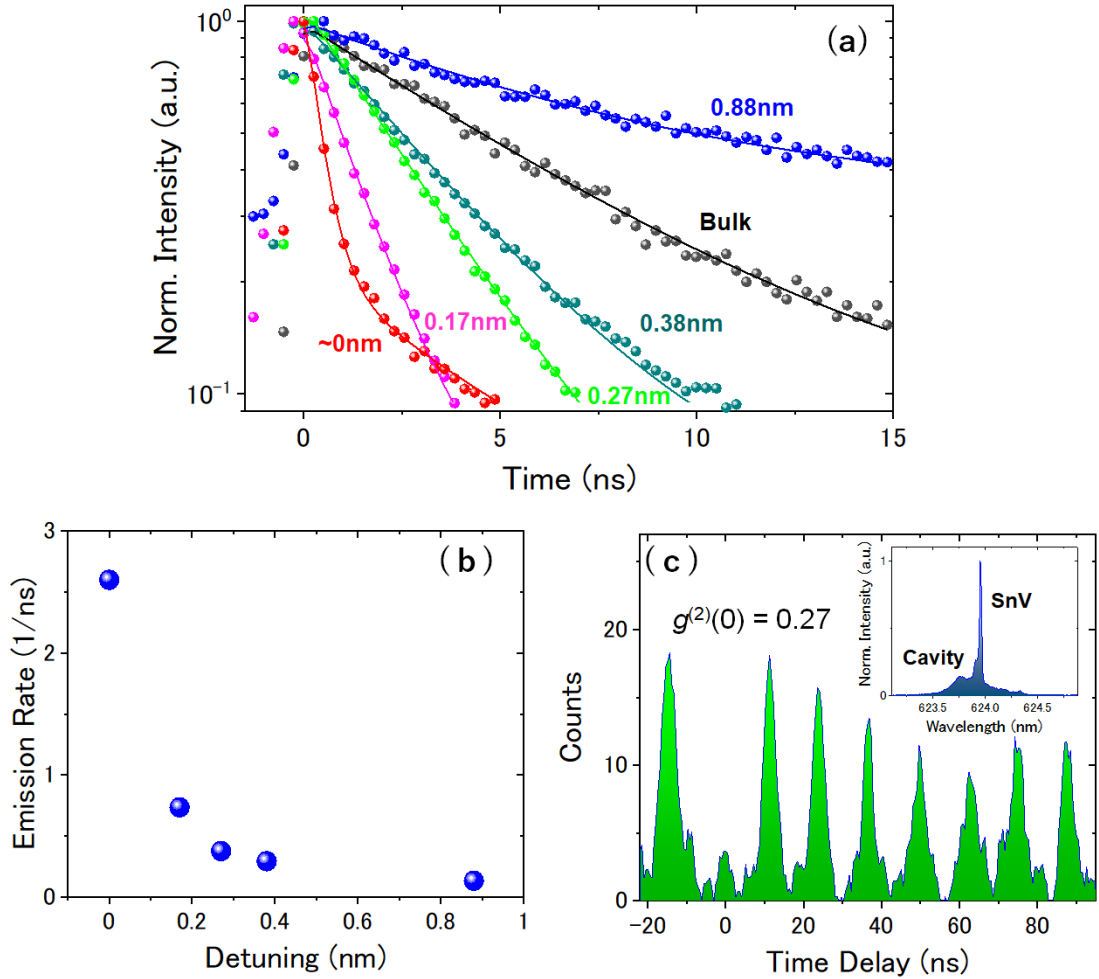

**Fig. 3.** (a) Time resolved PL spectra measured for the SnV under 5 different detuning conditions and for a SnV in bulk. The solid lines correspond to exponential fits convolved with the system's time resolution. (b) Extracted SnV emission rate as a function of detuning. (c) Second-order correlation $g^2(\tau)$ measured with a SnV-cavity detuning of $\Delta = 0.23$ nm. The inset shows the corresponding PL spectrum. All experimental data are recorded at 4 K.

In summary, we have demonstrated optical coupling between a single SnV center and a PhC nanobeam cavity. Using a cavity with a $Q$ factor of ~3,000, we observed clear intensity enhancement of the SnV emission resonantly coupled to the cavity by a factor of 12 and a 16-fold reduction in the SnV lifetime. Such a large enhancement of the SnV emission rate gives a Purcell factor for the ZPL of 37 and a high cavity coupling efficiency $\beta$ of 95%. These results prove the high potential of SnV centers for optical coupling to photonic nanocavities, which opens the door for a variety of diamond-based quantum photonic applications that require efficient and fast single photon emitters coupled to nanocavities [28]. We note that there is still room for further increase of the Purcell factor using cavity designs with extremely small $V$s and high $Q$s [38,39], as well as using precise position alignment techniques of single emitters [9,34,37]. We also note that the collection efficiency of the enhanced single photon emission can be largely improved using tapered waveguide for fiber coupling [42] and grating couplers [43]. In addition, our nanobeam design is compatible with strain-mediated tuning approaches of color center transitions. The incorporation of these functionalities into the demonstrated SnV-nanocavity coupled system is important for the development of spin-photon interfaces for future large-scale integration of color centers into integrated photonic circuits and quantum networks [44].

Note: During the process of finalizing this Letter, we became aware of a related work [45].

**Acknowledgments**

We would like to thank Bartholomeus Machielse, Prof. Amir Yacoby, Arjun Mirani, Smarak Maity and Linbo Shao for their helpful discussion. This work was supported by AFOSR (Grant No. FA9550-19-1-0376, and FA9550-20-1-0105), ARO MURI (Grant No. W911NF1810432), NSF RAISE TAQS (Grant No. ECCS-1838976), NSF STC (Grant No. DMR-1231319), NSF ERC (Grant No. EEC-1941583), DOE (Grant No. DE-SC0020376), DFG SFB 1375 "NOA" project C5, and ONR (Grant No. N00014-20-1-2425). K.K. acknowledges financial support from JSPS Overseas Research Fellowships (Project No. 202160592). B.P. acknowledges financial support through a Horizon 2020 Marie Sklodowska-Curie Actions global fellowship (COHESiV, Project No. 840968) from the European Commission. D.R. acknowledges support from the NSF GRFP and Ford Foundation fellowships. This work was performed in part at the Center for Nanoscale Systems (CNS), Harvard University.